\pdfoutput=1

\documentclass[11pt]{article}

\usepackage{EMNLP2023}

\usepackage{times}
\usepackage{latexsym}

\usepackage[T1]{fontenc}

\usepackage[utf8]{inputenc}

\usepackage{microtype}

\usepackage{inconsolata}

\usepackage{lipsum}
\usepackage{textcomp}  
\usepackage{scalerel} 
\usepackage{booktabs}
\usepackage{enumitem}
\usepackage{CJKutf8}
\usepackage{amssymb}

\usepackage[frozencache=true,cachedir=minted-cache]{minted}
\setminted{
frame=lines,
framesep=2mm,
baselinestretch=1.2,
fontsize=\footnotesize
}

\newcommand{\checkmrk}{\scalerel*{\includegraphics{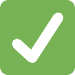}}{\textrm{\textbigcircle}}}
\newcommand{\cross}{\scalerel*{\includegraphics{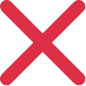}}{\textrm{\textbigcircle}}}
\newcommand{\kani}{{\fontfamily{qag}\selectfont
Kani}}
\newcommand{\crab}{\scalerel*{\includegraphics{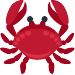}}{\textrm{\textbigcircle}}}
\newcommand{\kaniwithcrab}{\kani \hspace{1pt} \crab \hspace{1pt}}
\newcommand*\samethanks[1][\value{footnote}]{\footnotemark[#1]}
\newcommand{\inlinecode}[1]{\mintinline[fontsize=\normalsize]{python}{#1}}

\title{\kaniwithcrab: A Lightweight and Highly Hackable Framework for \\ Building Language Model Applications}

\author{Andrew Zhu\thanks{~~Equal contribution.}, \quad
  Liam Dugan\samethanks, \quad Alyssa Hwang, \quad
  Chris Callison-Burch \\
  University of Pennsylvania \\
  {\tt \{andrz,ldugan,ahwang16,ccb\}@seas.upenn.edu}
}

\begin{document}
\maketitle
\begin{abstract}
Language model applications are becoming increasingly popular and complex, often including features like tool usage and retrieval augmentation. However, existing frameworks for such applications are often opinionated, deciding for developers how their prompts ought to be formatted and imposing limitations on customizability and reproducibility. To solve this we present Kani: a lightweight, flexible, and model-agnostic open-source framework for building language model applications. Kani helps developers implement a variety of complex features by supporting the core building blocks of chat interaction: model interfacing, chat management, and robust function calling. All Kani core functions are easily overridable and well documented to empower developers to customize functionality for their own needs. Kani thus serves as a useful tool for researchers, hobbyists, and industry professionals alike to accelerate their development while retaining interoperability and fine-grained control.
\end{abstract}

\section{Introduction}
We introduce Kani, an open-source\footnote{Kani is available at \url{https://github.com/zhudotexe/kani}, free for use under the MIT license.} framework for building language model (LM) applications. Kani takes care of the basics of chat interaction---querying models, managing chat history, and calling external functions---allowing developers to write robust application code that is interoperable across any underlying language model. From this minimal base, developers can easily override the core features to implement more complex functionality like retrieval, web hosting, dynamic model routing, and tool usage tracking.

Unlike existing frameworks, Kani is lightweight and highly hackable, allowing developers to control their prompts, customize their models, and handle errors with ease. Our design philosophy is minimalist implementation with maximalist documentation: we implement a small number of universally useful core features while providing more complex application-specific examples in documentation. 

\begin{figure}
    \centering
    \includegraphics[width=\columnwidth]{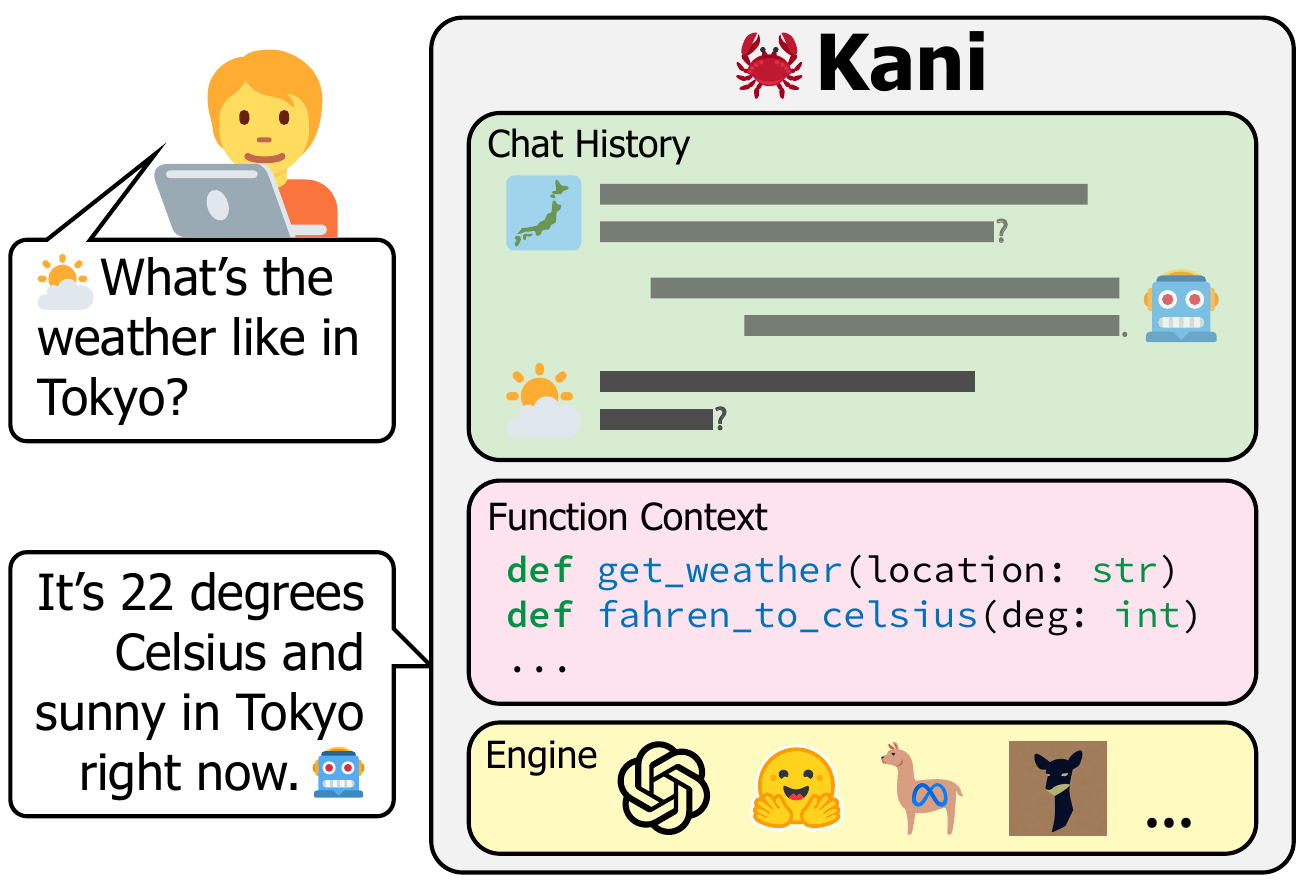}
    \caption{Kani is a \textit{lightweight} and \textit{flexible} framework that tracks chat history, calls inference engines, and manages callable functions in an un-opinionated manner---allowing researchers and developers to implement custom functionality easily and quickly.}
    \label{fig:overview}
\end{figure}

Kani is appealing to a wide range of developers. Hobbyists can get started with models like GPT-4, LLaMA v2, and Vicuna with as few as five lines of code. Industry professionals will enjoy the added robustness of automatic chat management and function retrying. Finally, researchers can appreciate the improved reproducibility afforded by fine-grained control over prompting. 

In this paper we provide a quick-start guide for developing with Kani (Section \ref{sec:quickstart}), an overview of our philosophy with comparisons to other frameworks (Sections \ref{sec:conceptual_overview}-\ref{sec:framework_comparison}), and a detailed tutorial on how to build more complex applications (Sections \ref{sec:developing}-\ref{sec:advanced}).

\section{Getting Started with Kani}
\label{sec:quickstart}
Let's start by discussing the basics of installing and querying language models with Kani. To start, Kani requires Python 3.10+ and is installed via pip.

\begin{minted}{bash}
$ pip install kani
\end{minted}

This command will install the core Kani dependencies. In order to use our pre-built engine classes for HuggingFace or OpenAI (Table \ref{tab:dependencies}), you must also include one or more ``extras'' with your pip installation command.

\begin{minted}{bash}
$ pip install kani[openai]
\end{minted}

In Figure \ref{fig:basic_example} we provide a minimal example to quickly get started with Kani in only five lines of code. We initialize the \inlinecode{OpenAIEngine} with our OpenAI API key, pass it into a new Kani object, and chat with the Kani using the built-in \inlinecode{chat_in_terminal()} function. With this, novice and advanced developers alike are able to easily query a variety of language models through Kani. 

\begin{table}
\small
\centering
\begin{tabular}{@{}l@{ }l@{ }l@{}}
\toprule
Platform & Engine & Extra    \\                 
\midrule
ChatGPT & \verb|OpenAIEngine| & \verb|openai| \\
GPT-4  & \verb|OpenAIEngine| & \verb|openai| \\
HuggingFace & \verb|HuggingEngine| & \verb|huggingface| \\ 
LLaMA v2 & \verb|LlamaEngine| & \verb|llama|\\ 
Vicuna v1.3 & \verb|VicunaEngine| & \verb|llama|\\
ctransformers & \verb|CTransformersEngine| & \verb|ctransformers| \\
LLaMA v2 &   \verb|LlamaCTransformersEngine|     & \verb|ctransformers|     \\
\bottomrule
\end{tabular}
\caption{The list of models and engines included in Kani with associated pip extras to add when installing. For example, to install Kani with support for HuggingFace Transformers, use \mintinline{text}{pip install 'kani[huggingface]'}.}
\label{tab:dependencies}
\end{table}

\begin{CJK}{UTF8}{min}
\section{Conceptual Overview}
\label{sec:conceptual_overview}
\subsection{What is the Kani object?}
The main atomic unit of processing in our framework is the titular Kani.\footnote{Kani (カニ) is Japanese for ``crab''. *\textit{snip snip}*} When developing applications with Kani you will mostly be spawning and manipulating different Kani objects. A Kani object consists of the following three parts:

\begin{enumerate}[noitemsep]
    \item \textbf{Inference Engine}: The underlying language model and associated framework.
    \item \textbf{Chat History}: The state of the conversation including system prompts.
    \item \textbf{Function Context}: The list of available callable functions, if any.
\end{enumerate}

To initialize a Kani all you need to pass in is an inference engine---the chat history will default to an empty list and callable functions are optional.
\end{CJK}

\begin{figure}
\begin{minted}{python}
from kani import Kani, chat_in_terminal
from kani.engines.openai import OpenAIEngine

engine = OpenAIEngine(api_key, model="gpt-4")
ai = Kani(engine)
chat_in_terminal(ai)
\end{minted}
\caption{A basic example showing how to initialize a Kani object and chat with GPT-4 \cite{openai2023gpt4} in only three lines of code.}
\label{fig:basic_example}
\end{figure}

\subsection{What does a Kani object do?}
When designing Kani, we wanted to implement the minimal set of features that allowed for the largest amount of flexibility and customization. Following this design principle, a Kani object does the following three things:

\begin{enumerate}[noitemsep]
    \item \textbf{Interfaces with Models}: Kani queries LMs via inference engines, allowing developers to swap models without editing the application.
    \item \textbf{Manages Chat History}: Kani tracks the token counts and turns of the conversation ensuring that models never exceed their context.
    \item \textbf{Exposes and Calls Functions}: Kani exposes functions to models, validates function calls, runs code, and returns output back to the inference engine. Kani also propagates all errors back to the model to allow for auto-retrying of failed function calls ensuring that such calls are robustly implemented.
\end{enumerate}

This core is flexible and minimalist, allowing for a wide array of emergent capabilities while simultaneously optimizing for robustness and scalability. For example, you can create a Kani that calls a retrieval function to augment its chat responses, following \citet{lewis-etal-2020-rag}, or a Kani that dynamically routes queries to different engines. All core functions of the Kani base class are asynchronous by default, allowing for easy web hosting and responsive applications.

\subsection{Where does Kani fit in the LM Stack?} 
\begin{figure}
    \centering
    \includegraphics[width=\columnwidth]{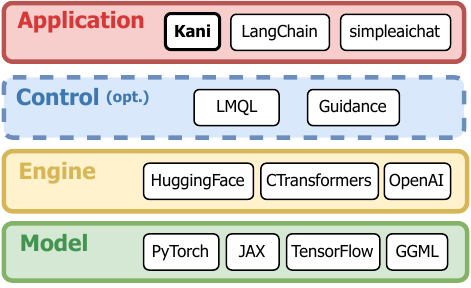}
    \caption{The different layers of the modern LM application stack. Kani sits at the Application layer and is simpler and more flexible than the competing frameworks. Additionally, Kani supports the usage of any lower level control or engine library, allowing developers to use their favorite frameworks alongside Kani.}
    \label{fig:layers}
\end{figure}

In Figure~\ref{fig:layers} we lay out our categorization of LM application libraries into four distinct layers: Model, Engine, Control, and Application. In this sub-section, we will give a brief overview of what each component of the stack accomplishes to help better contextualize how Kani fits in to the broader ecosystem of tools.

\paragraph{Model Layer.} In this layer, LM libraries assist with low-level procedures like matrix operations and hardware acceleration. Examples include PyTorch \cite{paszke2019pytorch}, TensorFlow \cite{abadi2016tensorflow}, and JAX \cite{jax2018github}. Kani is agnostic to the underlying model implementation so all Model libraries are compatible.

\paragraph{Engine Layer.} Libraries like HuggingFace \cite{wolf2020huggingfaces} and OpenAI \cite{chatgpt-etal-2022} in this layer manage elements of model inference such as sampling strategies and tokenization. Kani is interoperable across any Engine library by extending the \inlinecode{BaseEngine} class (see Section~\ref{sec:customization}). In an era characterized by an ever-changing state of the art, the ability to easily swap Engines without changing the application code is invaluable.

\paragraph{Control Layer.} Libraries in this optional layer handle complex control logic like dynamic prompt branching and tabular data prediction. Control libraries include LMQL \cite{Beurer_Kellner_2023} and Guidance \cite{lundberg2023Guidance}. Kani supports these libraries and can be configured to dynamically route queries on its own (see Section \ref{sec:sub_kani}), allowing for more robust inference.

\paragraph{Application Layer.} In the final layer, LM libraries provide the highest level of functionality by managing chat history, compiling prompts, creating function contexts, and handling errors. Examples of Application libraries include LangChain \cite{langchain2022}, simpleaichat \cite{simpleaichat2023}, and, of course, Kani. Kani provides a more flexible, interoperable, and streamlined experience to help any developer build LM applications.

\begin{figure*}
    \centering
    \includegraphics[width=\textwidth]{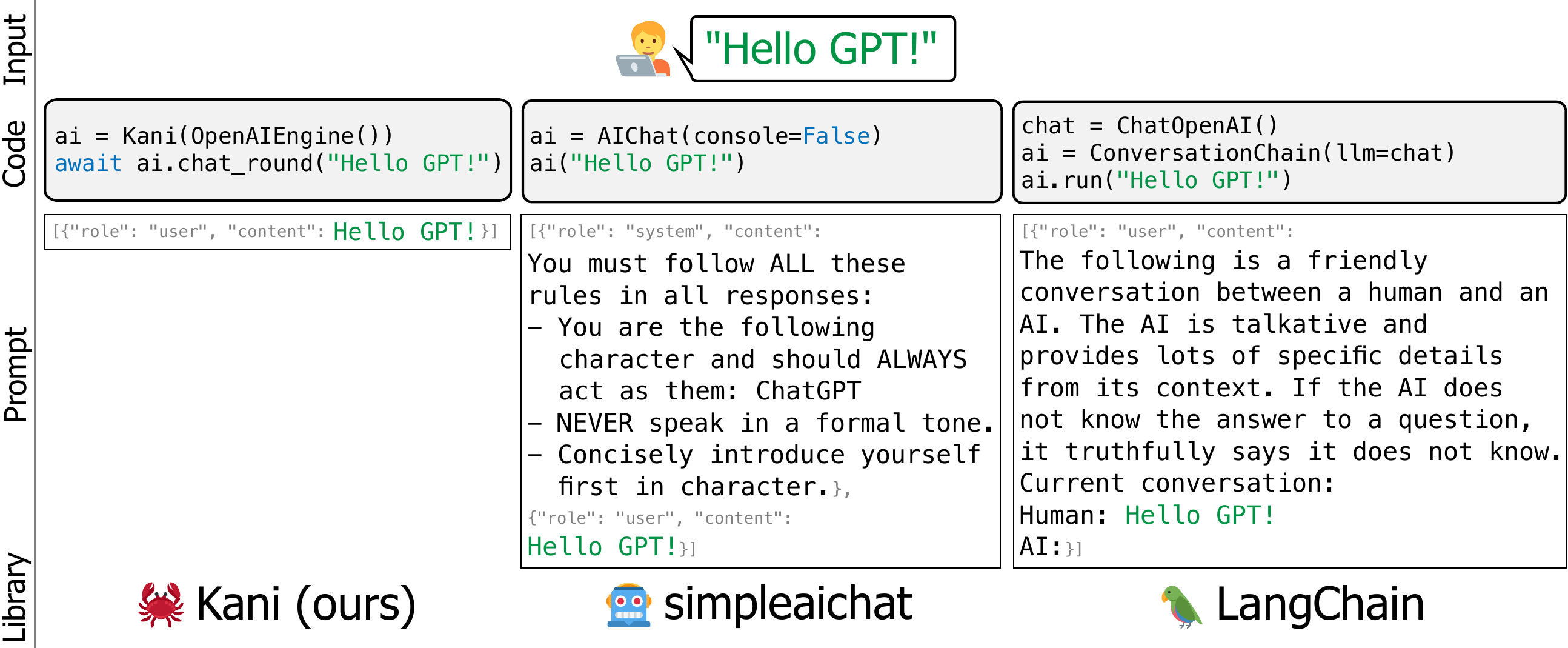}
    \caption{A comparison of prompting behavior between Kani and other competing frameworks. Kani does not edit developers' prompts under the hood in unexpected ways and allows for full control over what is passed to the model.}
    \label{fig:prompt_comparison}
\end{figure*}

\begin{table}
    \small
    \centering
    \begin{tabular}{l|ccc}
    \toprule
    &Kani&simpleaichat&LangChain\\
    \midrule
    Size (in MB)&13&26&156\\
    Dependencies&2&8&12\\
    \midrule
    Lightweight&\checkmrk&\checkmrk&\cross\\
    Chat Management&\checkmrk&\cross&\cross\\
    Function Retry&\checkmrk&\cross&\cross\\
    Model-Agnostic&\checkmrk&\cross&\checkmrk\\
    Un-opinionated&\checkmrk&\cross&\cross\\
    Extensive Docs&\checkmrk&\cross&\checkmrk\\
    \bottomrule
    \end{tabular}
    \caption{A feature comparison between Kani and competing frameworks. Kani is the only package that includes function retrying and chat management while still being lightweight and un-opinionated.}
    \label{tab:feature_comparison}
\end{table}

\section{Framework Comparison}
\label{sec:framework_comparison}
In this section, we compare Kani with simpleaichat \cite{simpleaichat2023} and LangChain \cite{langchain2022} to highlight Kani's strengths (see Table~\ref{tab:feature_comparison}).

\paragraph{Lightweight.} Kani is minimalist in both functionality and footprint: we implement essential features with fewer dependencies and less library-specific tooling while accomplishing more (see Table~\ref{tab:feature_comparison}). Paired with our detailed documentation, Kani's lean and efficient core of features allows developers to start easily and grow rapidly.

\paragraph{Chat History Management.} Unlike our contemporaries, Kani automatically tracks token counts and ensures that the maximum context length is never exceeded---letting developers focus on more exciting parts of their applications. Kani also lets developers easily customize this behavior by overriding \inlinecode{Kani.get_prompt()} (see Section \ref{sec:customize_prompt}).

\paragraph{Robust Function Calling.} In contrast to other frameworks, Kani \textit{guarantees} that function calls are valid by the time they reach developers' Python code. If a model calls a function incorrectly, Kani automatically provides feedback to the model and allows it to try again or follows developers' custom error handling (see Sections \ref{sec:retry_and_feedback} and \ref{sec:custom_error_handling}).

\paragraph{Model-Agnostic.} Kani provides a straightforward interface to use and interchange \textit{any} model. Developers can easily swap models without altering their source code, simplifying the process of switching models as newer ones are released.

\paragraph{Un-opinionated Prompting.} Unlike our contemporaries, Kani does not modify developers' prompts under the hood (see Figure \ref{fig:prompt_comparison}). We instead give developers full control to override and construct prompts themselves, leading to more robust, transparent, and reproducible source code.

\paragraph{Extensive Documentation.} Kani provides thorough and up-to-date documentation\footnote{\url{https://kani.readthedocs.io/}} on core library features with a particular focus on customizability. Our docs go beyond basic descriptions of features by including numerous examples of complex applications and guides on how to override and customize default behaviors.

\section{Developing Applications with Kani}
\label{sec:developing}
Now that we understand Kani's place in the broader ecosystem of tools, we will dive deeper into exactly how to develop LM applications with Kani.

\subsection{The Chat History}
Kani interacts with the user through \inlinecode{ChatMessage} objects, which are tracked in the chat history:

\begin{minted}{pycon}
>>> chat_in_terminal(ai, rounds=1)
USER: Hello Kani!
AI: Hello! How can I help?
>>> ai.chat_history
[ChatMessage(role=ChatRole.USER, 
             content="Hello Kani!"),
 ChatMessage(role=ChatRole.ASSISTANT, 
             content="Hello! How can I help?")]
\end{minted}

Following the OpenAI convention, each message contains the \inlinecode{role} (system, assistant, user, or function) and  \inlinecode{content} of the message.\footnote{Optionally, a user message can also contain a \mintinline{python}{name} (for multi-user conversations), and an assistant message can contain a \mintinline{python}{function_call} (discussed in Section \ref{sec:function_calling}).} Kani will pass in as much of this chat history as the engine's context window can hold as a default, which can be easily overridden (see Section \ref{sec:customization}). The chat history can also be saved or loaded in JSON format with \inlinecode{Kani.save()} and \inlinecode{Kani.load()} for ultimate control over the conversation context.

\subsection{Prompting}
Kani queries the underlying language model by providing a prompt, which is made of four parts:

\begin{enumerate}[noitemsep]
    \item \textbf{System Prompt} (optional): Content specifically for the \verb|system| role that typically defines high-level instructions for model responses.
    \item \textbf{Persistent Messages} (optional): Content that always appears at the top of the context window and will never be truncated.
    \item \textbf{Chat History}: The most recent messages that have not exceeded the context length.
    \item \textbf{User Message}: The current user input.
\end{enumerate}

The bulk of chat application interactions are a combination of these four components. For example, the system prompt can define a chatbot persona and the persistent messages can include a set of few-shot examples in the context (see Figure \ref{fig:few_shot}).

A system prompt and list of persistent messages can be passed into the Kani constructor at initialization: \inlinecode{Kani(engine, system='...',} \inlinecode{always_included_messages=[...])}. You can also define custom prompt behavior by overriding \inlinecode{Kani.get_prompt()} (see Section \ref{sec:customize_prompt}).

\begin{figure}
\begin{minted}{python}
shots = [ChatMessage.user("thank you"),
         ChatMessage.assistant("arigato"),
         ChatMessage.user("good morning"),
         ChatMessage.assistant("ohayo")]
ai = Kani(engine, always_included_messages=shots)
chat_in_terminal(ai)
# USER: crab
# AI: kani
\end{minted}
\caption{A basic example showing how to initialize a Kani with a few-shot prompt \cite{brown2020gpt3}. We can see that the Kani obeys the pattern and continues to translate English to Japanese in the chat session despite never being explicitly prompted to do so.}
\label{fig:few_shot}
\end{figure}

\subsection{Writing a Kani Application}
So far we have interacted with Kani exclusively through \inlinecode{chat_in_terminal()}. While this function is useful for testing, when building applications you may want to intercept the model output for logging, content filtering, or any other operation before serving it to the user. This can be done with \inlinecode{Kani.chat_round()}\footnote{\mintinline{python}{Kani.chat_round()} is an asynchronous method. This means that applications do not have to wait on it to finish and can instead perform other tasks while responses are being generated. To call these functions you must \mintinline{python}{await} them from an asynchronous context such as \mintinline{python}{asyncio.run()}.}, which executes one turn of the conversation and returns a \inlinecode{ChatMessage} from the system or assistant. We can then complete additional tasks and return the finalized response to the user, as demonstrated in Figure \ref{fig:async_example}.

\section{Function Calling}
\label{sec:function_calling}
Until this point, Kani objects had no abilities beyond text generation. Function calling (or ``tool usage'') makes Kani objects even more powerful as intelligent assistants.

\subsection{What is Function Calling?}
Function calling is the process of a model autonomously deciding to call a set of developer-defined functions. Models that have been fine-tuned to support function calling typically allow developers to provide function headers and docstrings in the prompt. When appropriate, the model will indicate that a certain function should be run with the given parameters in a JSON request. The developer then needs to receive this request, run the specified function with their own resources, and return the output back to the model. Without Kani, developers usually need to define and maintain their own logic to handle these requests.

Giving language models access to callable functions allows them to hook into various tools, like sending text messages, browsing the web, or creating a calendar event. Kani provides easy ways to document functions and handle errors, which let developers focus on writing full-featured applications without the fuss of tedious boilerplate.

\begin{figure}
\begin{minted}{python}
def is_toxic(message):
    # ... Run toxicity detection

async def chat_with_toxicity_filter(ai):
  while True:
    user_message = input("USER: ")
    message = await ai.chat_round(user_message)
    if is_toxic(message.content):
        message.content = "<Removed>"
    print("AI:", message.content)

ai = Kani(OpenAIEngine(api_key, model="gpt-4"))
asyncio.run(chat_with_toxicity_filter(ai))
\end{minted}
\caption{An example showing how to use Kani with additional output parsing. We query the engine using the \mintinline{python}{Kani.chat_round()} function and filter out toxic content.}
\label{fig:async_example}
\end{figure}

\subsection{Function Calling with Kani}
There are two ways to create a Kani with function calling capabilities. One way is to load them statically by making a subclass of the \inlinecode{Kani} base class and writing your functions as class methods with the \inlinecode{@ai_function()} decorator (see Figure \ref{fig:function_calling}). The other way to incorporate function calling is to load the functions dynamically by passing them in a list to the Kani constructor when instantiating a \inlinecode{Kani} base class or subclass (see Appendix \ref{app:dynamic_function_loading}). Querying a function-calling-enabled Kani is similar to what we have previously seen, except that \inlinecode{Kani.full_round()} should be used instead of \inlinecode{Kani.chat_round()}.\footnote{\url{https://kani.readthedocs.io/en/latest/api_reference.html\#kani.Kani.full_round}}

\begin{figure}
\begin{minted}{python}
class Unit(enum.Enum):
  FAHRENHEIT = "fahrenheit"
  CELSIUS = "celsius"
  
class WeatherKani(Kani):
  @ai_function()
  def get_weather(self, loc: Annotated[str, 
  AIParam(desc="The desired city")], unit: Unit):
    """Get the weather in a given location."""
    # ... Query some weather API
    return weather

chat_in_terminal(WeatherKani(engine))
# USER: What's the weather in San Francisco?
# AI: Thinking (get_weather)...
# AI: It's currently 72F in San Francisco.
\end{minted}
\caption{An example showing how to create a subclass of the base Kani and expose a function with \mintinline{python}{@ai_function}. Functions are given type annotations, triple-quoted docstrings, and \mintinline{python}{AIParam} descriptions to indicate to the model how they should be used.}
\label{fig:function_calling}
\end{figure}

\subsection{Documenting a Function}
Kani functions must be documented with native Python type annotations\footnote{We support primitive, compound, and enum Python types.} and docstrings (triple-quoted strings immediately following a function definition). You can optionally describe parameters even further by providing an \inlinecode{AIParam} annotation.

Proper function documentation not only helps language models use functions but also allows Kani to validate that a function is being called properly. For functions with proper type annotations, Kani \textit{guarantees} that all parameters are of the correct type before they reach your code. This feature is unique to Kani and allows for considerably more robust function calling.

\subsection{Retry \& Model Feedback}
\label{sec:retry_and_feedback}
When a function call returns an error, Kani will raise one of the following exception types:
\begin{itemize}[noitemsep]
    \item \textbf{NoSuchFunction}: The requested function was hallucinated and does not exist.
    \item \textbf{WrappedCallException}: The requested function raised an exception during execution.
    \item \textbf{TypeError}: The function exists, but the model hallucinated parameters that do not.
    \item \textbf{ValidationError}: The parameter names exist, but the model got the data types wrong.
\end{itemize}

If the model calls a function incorrectly, Kani will give it feedback by adding the error message to the chat history. This gives the model a chance to correct itself by retrying the call with new arguments or another function. Developers can customize the retry behavior or error messages with the \inlinecode{handle_function_call_exception()} method (see Section \ref{sec:custom_error_handling}). By anticipating common errors and automatically retrying function calls, Kani helps developers build more robust applications without the extra effort.

\section{Customization}
\label{sec:customization}
Kani is built on the philosophy that the developer should be in control of every aspect of an application. To accomplish this, Kani allows you to override and customize virtually all default behaviors of the library code. In this section we will briefly go over some common customizations developers may want to make.

\subsection{Customizing the Prompt}
\label{sec:customize_prompt}
Kani allows developers to control exactly what is being exposed to the language model by customizing the prompt builder. This can best be done by overriding the \inlinecode{Kani.get_prompt()} function.

\begin{figure}
\begin{minted}{python}
class AmnesiaKani(Kani):
  async def get_prompt(self):
    return self.always_included_messages 
      + self.chat_history[-2:]

chat_in_terminal(AmnesiaKani(engine))
# USER: Hi kani! My name is Andrew.
# AI: Hello Andrew! How can I assist you today?
# USER: What does "kani" mean in Japanese?
# AI: "Kani" in Japanese means "Crab".
# USER: What is my name?
# AI: As an AI, I don't have access to that data.
\end{minted}
\caption{A example showing how to customize the default \mintinline{python}{get_prompt()} function to only include the most recent two messages in the model prompt.}
\label{fig:amnesia_kani}
\end{figure}

In Figure \ref{fig:amnesia_kani} we show how you can customize the \inlinecode{Kani.get_prompt()} function to include only the most recent two messages, but this is just the tip of the iceberg. With custom prompt builders, developers can implement anything from dynamic prompt templating to fine-grained LMQL-style control prompts (see Appendix \ref{app:dynamic_prompt_templating}).

\subsection{Implementing a Custom Engine}
\label{sec:engines}
Kani interacts with language models through Engines. While Kani comes pre-packaged with a few starter engines, developers are encouraged to implement their own custom engines to adapt new language models or inference libraries for use with Kani. To create an engine, you must subclass the \inlinecode{BaseEngine} class. A new engine must implement:
\begin{enumerate}[noitemsep]
    \item \inlinecode{BaseEngine.message_len()}: Takes as input a \inlinecode{ChatMessage} and returns the token length of the message.
    \item \inlinecode{BaseEngine.predict()}: Takes in a list of \inlinecode{ChatMessage} and returns a new \inlinecode{Completion}.
    \item \inlinecode{BaseEngine.max_context_size}: Specifies the model's maximum token context length.
\end{enumerate}

Optionally, you can also choose to implement \inlinecode{BaseEngine.close()} to clean up resources or \inlinecode{BaseEngine.function_token_reserve()} if your engine needs to reserve some tokens for functions. Kani also comes with a few extra base classes and utilities to help you quickly build engines for models on HuggingFace \cite{wolf2020huggingfaces} (See Appendix \ref{app:engine_implementation}) or with an available HTTP API.\footnote{Built an engine for a popular model Kani doesn't support yet? Kani is open-source and greatly appreciates PRs with engine implementations for the latest models---see the \href{https://kani.readthedocs.io/en/latest/contributing.html}{contribution page} in our documentation.}

\begin{figure}
\begin{minted}{python}
class CustomExceptionKani(Kani):
  async def handle_function_call_exception(
    self, call, err, attempt):
    self.chat_history.append(ChatMessage.system(
      "The call encountered an error. Relay"
      f"it to the user sarcastically: {err}"))
    return attempt < self.retry_attempts

  @ai_function()
  def get_time(self):
    """Get the current time."""
    raise RuntimeError("The API is offline")

chat_in_terminal(CustomExceptionKani(engine))
# USER: What time is it?
# AI: Thinking (get_time)...
# AI: Well, it seems like our handy-dandy time
# API decided to take a coffee break...
\end{minted}
\caption{A example showing how to customize the \mintinline{python}{Kani.handle_function_call_exception()} function to return errors to the user in a sarcastic manner.}
\label{fig:sarcastic_kani}
\end{figure}

\subsection{Custom Error Handling}
\label{sec:custom_error_handling}
Kani calls \inlinecode{handle_function_call_exception()} whenever it encounters an error from a function. In Figure \ref{fig:sarcastic_kani}, we provide an example of overriding this function to tell our model to return function errors to the user in a sarcastic tone. While this is just a fun example, custom error messages can and often do serve a more utilitarian purpose by helping models retry functions more effectively.

\section{Advanced Usage}
\label{sec:advanced}
In this section, we'll look at some more advanced examples. For each of these use cases, we provide the full implementation in the GitHub repository.\footnote{\url{https://github.com/zhudotexe/kani/tree/main/examples}}

\begin{figure}
\begin{minted}{python}
class KaniWithSummary(Kani):
  @ai_function()
  async def summarize_conversation(self):
    """Get the summary of the conversation."""
    long_context_engine = OpenAIEngine(api_key, 
                          model="gpt-4-32k")
    sub_kani = Kani(long_context_engine, 
        chat_history=self.chat_history[:-2])
    summary = await sub_kani.chat_round(
    "Please summarize the conversation so far.")
    return summary.content

chat_in_terminal(KaniWithSummary(engine))
# USER: Tell me about trains.
# AI: Trains are modes of long-distance transport
# [Multiple turns of conversation...]
# USER: Summarize the conversation.
# AI: Thinking (summarize_conversation)...
# AI: Our chat began with a general overview 
# about trains and how railway systems work...
\end{minted}
\caption{A example showing how to use sub-kani spawning to dynamically resize the context window of the model depending on a user query. Note that the base \mintinline{python}{"gpt-4"} kani spawns a \mintinline{python}{"gpt-4-32k"} sub-kani in order to capture the full conversation for summarization.}
\label{fig:sub_kani}
\end{figure}

\subsection{Sub-Kanis}
\label{sec:sub_kani}
When used in conjunction with function calling, Kani can choose to spawn ``sub-Kani''---self-contained ``agents'' capable of performing their own tasks then reporting to the parent with their results.

For example, you might have the parent Kani use a cheaper, faster model with a smaller context length. If you need it to perform a task that requires more context, you can spawn a sub-Kani using a more expensive, slower model with a larger context. In Figure \ref{fig:sub_kani}, we show how you can spawn a sub-Kani inside a callable function and copy the chat history to accomplish this.

Of course, the sub-Kani you spawn doesn't have to be a vanilla Kani---you could imagine having multiple different Kani types with different sets of functions or engines, each capable of performing their own specialized tasks.

\subsection{Retrieval}
Language models can be augmented with an external factual database that they can retrieve information from, allowing them to access more relevant and up-to-date information without having to re-train on more recent events.

In Figure \ref{fig:retrieval}, we demonstrate how Kani's function calling can be used to retrieve information from a data source like Wikipedia. Since retrieved articles might be longer than the model's maximum context window, you may want to combine this with the previous summarization example for maximum efficacy.

\begin{figure}
\begin{minted}{python}
class WikipediaKani(Kani):
  @ai_function()
  async def wikipedia(self, title: Annotated[
      str, AIParam(desc='The article title')]):
    """Get information from Wikipedia."""
    if page := await wikipedia_client.get(title):
      return page
    return f"Page {title!r} does not exist"

  @ai_function()
  async def search(self, query: str):
    """Find article titles given a query."""
    titles = await wikipedia_client.search(query)
    return json.dumps(titles)

chat_in_terminal(WikipediaKani(engine))
# USER: Tell me about the Tokyo Yamanote line.
# AI: Thinking (search)...
# AI: Thinking (wikipedia)...
# AI: The Yamanote is a loop service in Tokyo...
\end{minted}
\caption{A example showing how to make a retrieval agent in Kani using custom AI function declarations. The WikipediaKani exposes the two functions (\mintinline{python}{search()} and \mintinline{python}{wikipedia()}) to the model which then calls both in order to retrieve the page for generation.}
\label{fig:retrieval}
\end{figure}

\subsection{Hosting a Kani Online}
What if you want to host a web service to allow users to chat with a Kani online? In Figure \ref{fig:web_kani}, we show how you can host and connect to a Kani on a webserver using a WebSocket connection. 

\begin{figure}
\begin{minted}{python}
engine = OpenAIEngine(api_key, model="gpt-4")
app = FastAPI()

@app.websocket("/chat")
async def kani_chat(websocket: WebSocket):
  await websocket.accept()
  ai = Kani(engine)
  while True:
    data = await websocket.receive_text()
    resp = await ai.chat_round(data)
    await websocket.send_text(resp.content)
\end{minted}
\caption{A example showing how to host and query Kani on the web using FastAPI and WebSockets.}
\label{fig:web_kani}
\end{figure}

We use FastAPI\footnote{\url{https://fastapi.tiangolo.com/}} to run this webserver. To connect to our server, we can use any client that supports WebSockets, like Insomnia.\footnote{\url{https://insomnia.rest/}} Web frameworks like FastAPI and Flask 2 allow route methods to be asynchronous, meaning you can await a Kani method from within your route method without needing to call \inlinecode{asyncio.run()}.

\section{Conclusion}
In this paper we presented Kani, a lightweight and highly customizable framework for building chat applications. At its core, Kani lets developers use the same application code across all language model backends and robustly implements convenient quality-of-life features like chat history management, function validation, and error handling.

We believe that the design of our tools is as important as the tools themselves. Well-designed tools impose far less friction when using them, freeing up developers' hands from fighting bugs and racking up tech debt. This is especially important now that the LLM landscape is so turbulent with new and improved models being released more often than ever. Kani offloads the burden of tedious language model management without locking developers into onerous default paradigms, giving developers back control over their applications---hopefully making the landscape a bit less turbulent.

\section*{Limitations}
One limitation of the Kani framework is that not all models are natively chat models. Given our design decision to maintain the internal state as a chat, with such attributes as roles and system prompts present, implementing interfaces for traditional completion-based language models is more difficult than it otherwise could have been with a different internal memory organization scheme.

Another limitation of our work is the lack of native function calling support in all models, making our defining features around robust function calling irrelevant for such models (e.g. LLaMA). However, the customizable nature of Kani allows developers who want such a feature to simply create a new Engine class and implement custom output parsing logic to recognize and route function calls themselves. Kani thus gives developers maximum flexibility in the creation of their applications.

\section*{Acknowledgments}
We would like to thank the members of the lab of Chris Callison-Burch for their testing and detailed feedback on the contents of both this paper and the Kani repository. In addition, we'd like to thank Henry Zhu (no relation to the first author) for his early and enthusiastic support of the project.

This research is based upon work supported in part by the IARPA HIATUS Program (contract 2022-22072200005), and the NSF (Award 1928631). Approved for Public Release, Distribution Unlimited. The views and conclusions contained herein are those of the authors and should not be interpreted as necessarily representing the official policies, either expressed or implied, of IARPA, NSF, or the U.S. Government.

\bibliography{kani}
\bibliographystyle{acl_natbib}

\appendix
\section{Size Measurement}
This section describes how we measured the number of dependencies of a library and its size in the framework comparison table (Table \ref{tab:feature_comparison}). We define a library's dependency count as the number of top-level dependencies that are installed when installing the library from pip without any extras. We measure the size of a library by installing it in a fresh Python virtual environment, running a command to measure the size of installed packages, and removing the size of the \inlinecode{pip} and \inlinecode{setuptools} packages (packaging utilities included in every Python environment). Specifically, we used the following shell commands:

\begin{minted}{text}
python3.10 -m venv venv
source venv/bin/activate
pip install {kani|simpleaichat|langchain}
du -h venv/lib/python3.10/site-packages
\end{minted}

\section{Dynamic Prompt Templating}
\label{app:dynamic_prompt_templating}
Below is an example of dynamically customizing a system prompt to include the phrase ``Always act like \texttt{<persona>}'' if a user types a chat message containing the phrase ``act like.'' This is a flexible alternative to hard-coding persona logic as is common in other repositories.

\begin{minted}{python}
class PersonaKani(Kani):
  def get_persona_prompt(self):
    if self.persona:
        return ChatMessage.system(
             f"Always act like {self.persona}.")

  async def get_prompt(self):
    prev = self.chat_history[-1].content
    if match := re.search(r"act like (.+)", prev):
      self.persona = match[1]
    return [self.get_persona_prompt()] + 
          await super().get_prompt()
\end{minted}

\section{Tracking Function Calls}
\label{app:track_function_calls}
Below we show an example of overriding the default \inlinecode{do_function_call()} method to additionally keep track of how many times a model called a function and how often it was successful.

\begin{minted}{python}
class TrackCallsKani(Kani):
  def __init__(self, *args, **kwargs):
    super().__init__(*args, **kwargs)
    self.successful_calls = collections.Counter()
    self.failed_calls = collections.Counter()

  async def do_function_call(self, call):
    try:
      res = await super().do_function_call(call)
      self.successful_calls[call.name] += 1
      return res
    except FunctionCallException:
      self.failed_calls[call.name] += 1
      raise

  @ai_function()
  def get_time(self):
    """Get the current time."""
    raise RuntimeError("The time API is offline")

  @ai_function()
  def get_date_and_time(self):
    """Get the current day and time."""
    return str(datetime.datetime.now())
\end{minted}

After chatting with our Kani, we can print out the new \inlinecode{successful_calls} and \inlinecode{failed_calls} variables to recover statistics on how well our models are calling our custom AI functions.
\begin{minted}{pycon}
>>> chat_in_terminal(TrackCallsKani(engine))
USER: What time is it?
AI: Thinking (get_time)...
AI: Thinking (get_date_and_time)...
AI: The current time is 22:42.
>>> ai.successful_calls
Counter({'get_date_and_time': 1})
>>> ai.failed_calls
Counter({'get_time': 1})
\end{minted}

This behavior is particularly useful for researchers studying language model tool usage and similar customizations can be easily made to other core functions to add more tracking.

\section{Dynamic Function Loading}
\label{app:dynamic_function_loading}
Rather than statically defining the list of functions a Kani can use in a class, you can also pass a list of functions to the Kani constructor when you initialize a Kani. To do this we need to use the special \inlinecode{Kani.AIFunction} class (which is similar to the traditional \inlinecode{@ai_function} decorator).

\begin{minted}{python}
def my_cool_function(foo: str,
  bar: Annotated[int, AIParam(desc="Cool int")]):
  """Do some cool things."""

engine = OpenAIEngine(api_key, model="gpt-4")
functions = [AIFunction(my_cool_function)]
ai = Kani(engine, functions=functions)
\end{minted}

This is particularly useful when spawning sub-Kani, as such agents can be dynamically given only a particular subset of the functions defined in the parent to help increase function call accuracy.

\section{Example Engine Implementations}
\label{app:engine_implementation}
In this section, we include the HuggingFace \cite{wolf2020huggingfaces} and LLaMA v2 \cite{touvron2023llama} engine implementations to demonstrate how a developer might implement new engines. The HuggingFace engine acts as a base engine class that implements common logic for all HuggingFace models, while the LLaMA v2 engine extends the base HuggingFace class with the model-specific prompt and delimiting tokens.

\begin{figure*}
\begin{minted}{python}
class HuggingEngine(BaseEngine, abc.ABC):
  def __init__(
    self,
    model_id: str,
    max_context_size: int,
    device: str | None = None,
    tokenizer_kwargs: dict = {},
    model_load_kwargs: dict = {},
    **hyperparams,
  ):
    self.model_id = model_id
    self.max_context_size = max_context_size
    self.tokenizer = AutoTokenizer.from_pretrained(model_id, **tokenizer_kwargs)
    self.model = AutoModelForCausalLM.from_pretrained(model_id, **model_load_kwargs)
    self.hyperparams = hyperparams

    if device is None:
      device = "cuda" if torch.has_cuda else "cpu"
    self.device = device
    if self.model.device.type != self.device:
      self.model.to(device)
      
  @abc.abstractmethod
  def build_prompt(
    self, messages: list[ChatMessage], functions: list[AIFunction] | None = None
  ) -> str | torch.Tensor:
    """Given the list of messages from kani, build either a single string
    representing the prompt for the model, or build the token tensor."""
    raise NotImplementedError

  async def predict(
    self, messages: list[ChatMessage], functions: list[AIFunction] | None = None, **hyperparams
  ) -> Completion:
    """Given the current context of messages and available functions, get the next
    predicted chat message from the LM."""
    prompt = self.build_prompt(messages, functions)
    if isinstance(prompt, str):
      tokenized = self.tokenizer(prompt, return_tensors="pt", return_length=True)
      input_len = int(tokenized.length)
      input_toks = tokenized.input_ids
    elif isinstance(prompt, torch.Tensor):
      input_toks = prompt
      input_len = len(input_toks[0])
    else:
      raise TypeError("build_prompt should either return a str or a Tensor.")
    # move the input tensor to the right device
    if input_toks.device.type != self.device:
      input_toks = input_toks.to(self.device)
    # set up hyperparams for HF decode
    hyperparams = {**self.hyperparams, **hyperparams}
    # run it through the model
    output = self.model.generate(input_toks, **hyperparams)
    # decode to tokens
    # the completion shouldn't include the prompt or stop token
    content = self.tokenizer.decode(output[0][input_len:-1]).strip()
    return Completion(ChatMessage.assistant(content), prompt_tokens=input_len,
              completion_tokens=len(output[0]) - (input_len + 1))
\end{minted}
\end{figure*}

\begin{figure*}
\begin{minted}{python}
class LlamaEngine(HuggingEngine):
  def __init__(self, model_id: str = "meta-llama/Llama-2-7b-chat-hf", *args, **kwargs):
    kwargs.setdefault("max_context_size", 4096)  # LLaMA has 4096 token window
    super().__init__(model_id, *args, **kwargs)

  def build_prompt(self, messages: list[ChatMessage], functions: list[AIFunction] | None = None):
    tokens = []
    prompt_buf = []  # parts of the user-assistant pair
    for message in messages:
      if message.role == ChatRole.USER:
        prompt_buf.append(f"{B_INST} {message.content} {E_INST}")
      elif message.role == ChatRole.ASSISTANT:
        prompt_buf.append(f" {message.content} ")
        # turn the current round into tokens
        prompt_round = "".join(prompt_buf)
        # if we see a " {E_INST}{B_INST} " we should replace it with empty string
        # (it happens immediately after a system + user message)
        prompt_round.replace(f" {E_INST}{B_INST} ", "")
        tokens.extend(self.tokenizer(prompt_round))
        # tokenizer adds the BOS token but not the EOS token
        tokens.append(eos_token_id)
        prompt_buf.clear()
      else:
        prompt_buf.append(f"{B_INST} {B_SYS}{message.content}{E_SYS} {E_INST}")
    # flush rest of prompt buffer (probably a user message) into tokens
    if prompt_buf:
      tokens.extend(self.tokenizer("".join(prompt_buf)))
    return torch.tensor([tokens], device=self.device)

  def message_len(self, message: ChatMessage) -> int:
    if message.role == ChatRole.USER:
      # <s> [INST] {} [/INST] -> 7
      return self.tokenizer(message.content, return_length=True).length[0] + 7
    elif message.role == ChatRole.ASSISTANT:
      # {} </s> -> 2
      return self.tokenizer(f" {message.content} ", return_length=True).length[0] + 2
    # <s> [INST] <<SYS>>\n{}\n<</SYS>>\n\n [/INST] -> 20
    return self.tokenizer(message.content, return_length=True).length[0] + 20
\end{minted}
\end{figure*}
\end{document}